\documentclass[aps,twocolumn,groupedaddress,nopacs,amsmath]{revtex4}

\usepackage{epsf,latexsym}
\usepackage{amssymb,amsmath}
\usepackage{times}

\newcommand{\ket}[1]{| #1 \rangle}

\newcommand{\ignore}[1]{}

\newcommand{\be}{\begin{equation}}
\newcommand{\ee}{\end{equation}}
\newcommand{\ba}{\begin{eqnarray}}
\newcommand{\ea}{\end{eqnarray}}

\def\CC{{\rm\kern.24em \vrule width.04em height1.46ex depth-.07ex
    \kern-.30em C}}
\def\P{{\rm I\kern-.25em P}}

\def\RR{{\rm
         \vrule width.04em height1.58ex depth-.0ex
         \kern-.04em R}}

\def\bbbc{{\mathchoice {\setbox0=\hbox{$\displaystyle\rm C$}\hbox{\hbox
to0pt{\kern0.4\wd0\vrule height0.9\ht0\hss}\box0}}
{\setbox0=\hbox{$\textstyle\rm C$}\hbox{\hbox
to0pt{\kern0.4\wd0\vrule height0.9\ht0\hss}\box0}}
{\setbox0=\hbox{$\scriptstyle\rm C$}\hbox{\hbox
to0pt{\kern0.4\wd0\vrule height0.9\ht0\hss}\box0}}
{\setbox0=\hbox{$\scriptscriptstyle\rm C$}\hbox{\hbox
to0pt{\kern0.4\wd0\vrule height0.9\ht0\hss}\box0}}}}

\def\bbbz{{\mathchoice {\hbox{$\sf\textstyle Z\kern-0.4em Z$}}
{\hbox{$\sf\textstyle Z\kern-0.4em Z$}}
{\hbox{$\sf\scriptstyle Z\kern-0.3em Z$}}
{\hbox{$\sf\scriptscriptstyle Z\kern-0.2em Z$}}}}

\newcommand{\putfig}[2]{$$\leavevmode\hbox{\epsfxsize=#2 cm
   \epsffile{#1.eps}}$$}

\begin{document}

\title{Beam-splitters don't have memory: a comment on ``Event-based corpuscular model for quantum optics experiments'' by K.Michielsen {\em et al.}}

\author{Radu Ionicioiu}
\affiliation{Centre for Quantum Computer Technology, Macquarie University, Sydney NSW 2109, Australia}

\begin{abstract}
In a recent article (arXiv:1006.1728) K.Michielsen {\em et al.} claim that a simple corpuscular model can explain many quantum optics experiments. We discuss these claims and show that their proposal fails at several levels. Finally, we propose an experiment to falsify the model.
\end{abstract}

\maketitle

The proposal we discuss here has been introduced recently in several articles \cite{ebcm,mzi,wdc}. This model, called the event-based corpuscular model (EBCM), claims to explain many quantum optics experiments, including interference, Einstein-Podolsky-Rosen and Hanbury Brown-Twiss effects. The model uses a classical simulation on a event-by-event basis, is causal and satisfies Einstein's locality. Moreover, the authors claim that EBCM is universal and can accommodate even correlations not explained by Maxwell's theory \cite{universal}. Here we critically evaluate these claims and show that EBCM fails at several levels as a model of reality.

\subsection{A logical problem}

Reading section II.3 in Ref.\cite{ebcm} it becomes clear that the authors do not accept the existence of destructive interference. Mathematically, they refuse the reality of two terms (events) with zero sum (annihilating each-other) \cite{ebcm}: ``Therefore, if we want to identify these ÒeventsÓ with the clicks that we observe, we run into a logical contradiction: To perform the sums in Eq. (3), we have to generate events that in the end cannot be interpreted as clicks since in this particular case no detector clicks are observed."

Extending this type of argument, the authors should also deny the existence of, e.g., two equally opposing forces acting on a classical object. Since the total force is zero, the two forces cancelling each-other cannot have a ``real existence'' as they produce no observable effect (i.e., movement).

\subsection{Failure as a corpuscular model}

The first corpuscular theory has been proposed by Newton following his optics experiments. As the theory failed to explain an accumulating body of data, it has been gradually abandoned in the favour of the wave model. We give a short list of phenomena not explained in a corpuscular model.

\noindent{\bf Diffraction.} A corpuscular theory doesn't explain the broadening of a plane wave passing through a narrow slit (with dimensions of the same order as the wavelength). This is also valid in the EBCM model and the authors do not offer any explanation of diffraction.

\noindent{\bf Resonant cavities and antennas.} Since a corpuscular theory has no concept of wavelength, there is no explanation of how a resonant cavity or an antenna works.

\noindent{\bf Photon bunching.} Due to their bosonic nature, photons exhibit bunching at a beam-splitter when entering from different ports. Thus the state $\ket{1}_a\ket{1}_b$ evolves to $\ket{2}_a\ket{0}_b+ \ket{0}_a\ket{2}_b$, where the subscripts $a,b$ denote the corresponding input or output modes. Such a behaviour is not only unexplained, but falls completely outside the model. In EBCM only one photon at a time is allowed, by construction, to enter the beamsplitter \cite{ebcm}; therefore it cannot describe multi-photon states like the recently observed $NOON$ states $\ket{N}_a\ket{0}_b+ \ket{0}_a\ket{N}_b$ \cite{noon}.

As a corpuscular model claiming to be universal \cite{universal}, EBCM inherits the same problem of explaining the above phenomena supported by hundreds of experiments. However, in several articles published about the EBCM the authors are silent about these facts \cite{ebcm,mzi,wdc}.

\subsection{Failure to explain all the experimental data}

The authors acknowledge \cite{ebcm} that EBCM does not faithfully reproduce {\em all} experimental data -- there is a transient period of a few hundred events in which the model fails to give the observed statistics. The first few hundred events are in stark contrast with the experimental data, contradicting not only the average number $\overline N $ of detector clicks, but also the standard deviation $\Delta N$ (a point not discussed by the authors). For a perfectly balanced Mach-Zehnder interferometer (MZI) in which detector $D_0$ should register no clicks, during the transient period the simulation shows instead a large number (the majority) of events exactly in detector $D_0$. Such fluctuations are too large to be explained statistically -- on average, the standard deviation should be $\sim \sqrt N$, where $N$ is the number or events. This bound is clearly violated in the simulations during the initial transient period. A physicist comparing experimental data with the predictions of the theory will rightly conclude that the model fails to reproduce the experiments. Instead of offering an explanation, the authors declare lightly \cite{ebcm}: ``However, as ample numerical simulations demonstrate, for many events, these first few "wrong" events do not significantly contribute to the averages and are therefore irrelevant for the comparison of the event-based simulation results with those of a wave theory.''

It is not a standard scientific practice to exclude a data set unexplained by theory, even if these are only ``a few hundred'' \cite{ebcm} -- what if the experimental data are only a few hundred measurements? This observation raises questions about the ability of EBCM to explain the experiments.

\subsection{Failure to faithfully model a real beamsplitter}

A central element of EBCM is the beamsplitter, modelled as a deterministic learning machine (DLM). The DLM has a set of internal registers (memory) which are continuously updated according to the state of the incident photon. A natural question is: Does a DLM faithfully model a real beamsplitter? Two arguments show that it fails to do so:\\
{\bf Input/output symmetry.} A real beamsplitter is symmetric in inputs/outputs. In the laboratory one can rotate a BS by 180 degrees and swap inputs with outputs without affecting the experiments. This is not possible with the DLM model of a beamsplitter (see Fig.1 in \cite{mzi}), where the outputs cannot be used as inputs.\\
{\bf Multi-photon input.} The DLM model of a beamsplitter also fails to explain what happens when more that one photon is present at the input. The authors explicitly state \cite{wdc}: ``The DLM receives a message on either input channel 0 or 1, never on both channels simultaneously.''

\begin{figure}[t]
\putfig{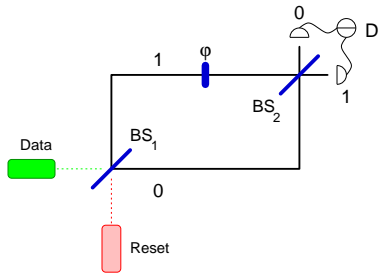}{5}
\caption{A resetting procedure for DLM (beamsplitter) memory. The data source (green) emits photons whose statistics one wishes to measure. At certain times the reset source (red) emits photons on input 1 of the BS in order to erase the memory of the device.}
\label{reset}
\end{figure}

\subsection{A gedankenexperiment to falsify EBCM}

Next we propose an experiment to falsify the predictions of EBCM. The main difference between a standard wave theory (WT) -- classical or quantum -- and EBCM is in the way it treats single events. In WT a single event (i.e., a photon entering the interferometer) leaves no trace (memory) in the apparatus. The device behaves in the same way for the first event and for event number 10000. This is completely different in the EBCM, where the beamsplitter/DLM has a memory and the final statistics is build gradually according to the past events. This explains the existence of the transient period with the ``wrong'' statistics. The main idea behind the proposed experiment is to make the transient period arbitrarily large, such that the device cannot reproduce the observed measurements.

In WT is irrelevant if we collect the full data set in one big run or in several smaller ones. This is a standard practice in the lab, where an experiment can last several weeks and data collection is segmented in smaller runs of several hours. It also does not matter if in between those runs we shine or not light on the interferometer, as long as this light is not collected (the detectors are shut down). However, this is not so in EBCM.

The experiment is sketched in Fig.\ref{reset}. A data source (green) sends photons (``messages'', in EBCM) to the 0-input of a standard MZI (Fig.\ref{reset}). We collect a block of data within the transient period, after which we ``reset'' the beamsplitter, {\em while keeping the source and detectors unchanged}. One way of doing a (software) reset is to use another source (the reset source, red) to send photons to the 1-input of the MZI. The role of the second source is to modify/reset the internal registers of the DLM and consequently to change the statistics of the data photons. As we are interested only in the statistics of the data source, we do not register the reset photons (e.g., we switch off the detectors during the time the reset source operates).

If each data block is collected after the DLM/beamsplitter has been reinitialised, the (arbitrarily large) set of data will not converge to the expected averages. Instead it will show the same mean value $\overline N$ and standard deviation $\Delta N$ as the transient period which, as noted in \cite{ebcm}, do dot reproduce the experiments. In WT, since the device has no memory of photons passing through it, intercalating reset photons (not detected) with data photons does not change the statistics.

Many variations of this scenario are experimentally possible. We can have an alternating sequence of data and reset photons $(N_1, N'_1, N_2, N'_2, \ldots)$, where $N_i\, (N'_i)$ denotes the number of data (reset) photons. In one experiment we can send a reset photon after each data photon; in a different one we can send $10^4$ reset photons after 100 data photons. Provided the sum of all data photons is the same $\sum_i N_i= N$, a standard wave theory predicts that all these different experiments will show the same statistics. In contrast, in EBCM all these experiments will have (in general) different statistics.

Another way to reset the memory of the DLM/beamsplitter is to replace it with a new one (having the internal registers reinitialised) and then collect another batch of data. This ``hardware'' reset is more challenging experimentally, but theoretical possible. Again, in a standard WT this will not affect the data, in contrast to the EBCM, where it resets the statistics back to the transient period.

In conclusion, we have shown that the EBCM fails at several levels to describe a large body of phenomena and as such cannot be considered a valid model of reality.



\begin{thebibliography}{}
\bibitem{universal} ``The event-based corpuscular model (EBCM) that we introduce in this paper is universal in that it can, without modification, be used to explain why photons build up interference patterns, why they can exhibit correlations that cannot be explained within MaxwellÕs theory and so on.'' \cite{ebcm}
\end{thebibliography}
\end{document}